\documentclass[aps,prb,twocolumn,floats,showpacs]{revtex4}
\usepackage{epsfig}
\usepackage{color}
\usepackage{bm}
\usepackage{latexsym}

\begin{document}
\newcommand{\hide}[1]{}
\newcommand{\tbox}[1]{\mbox{\tiny #1}}
\newcommand{\half}{\mbox{\small $\frac{1}{2}$}}
\newcommand{\sinc}{\mbox{sinc}}
\newcommand{\const}{\mbox{const}}
\newcommand{\trc}{\mbox{trace}}
\newcommand{\intt}{\int\!\!\!\!\int }
\newcommand{\ointt}{\int\!\!\!\!\int\!\!\!\!\!\circ\ }
\newcommand{\eexp}{\mbox{e}^}
\newcommand{\bra}{\left\langle}
\newcommand{\ket}{\right\rangle}
\newcommand{\EPS} {\mbox{\LARGE $\epsilon$}}
\newcommand{\ar}{\mathsf r}
\newcommand{\im}{\mbox{Im}}
\newcommand{\re}{\mbox{Re}}
\newcommand{\bmsf}[1]{\bm{\mathsf{#1}}}
\newcommand{\mpg}[2][1.0\hsize]{\begin{minipage}[b]{#1}{#2}\end{minipage}}

\title{On the generalized dimensions of multifractal eigenstates}

\author{J. A. M\'endez-Berm\'udez and A. Alcazar-L\'opez}
\affiliation{Instituto de F\'{\i}sica, Benem\'erita Universidad Aut\'onoma de Puebla, Apartado Postal J-48, Puebla 72570, Mexico}
\author{Imre Varga}
\affiliation{Elm\'eleti Fizika Tansz\'ek, Fizikai Int\'ezet, Budapesti M\H uszaki \'es Gazdas\'agtudom\'anyi Egyetem, H-1521 Budapest, Hungary}

\date{\today}

\begin{abstract}
Recently, based on heuristic arguments, it was conjectured that an intimate
relation exists between any multifractal dimensions, $D_q$ and $D_{q'}$, of the eigenstates of critical random matrix ensembles:
$D_{q'} \approx qD_q[q'+(q-q')D_q]^{-1}$, $1\le q, q' \le 2$. Here, we verify this relation by extensive numerical calculations 
on critical random matrix ensembles and extend its applicability to $q<1/2$ but also to deterministic models producing multifractal 
eigenstates and to generic multifractal structures. We also demonstrate, for the scattering version of the power-law banded
random matrix model at criticality, that the scaling exponents $\sigma_q$ of the inverse moments of Wigner delay times,
$\bra \tau_{\tbox W}^{-q} \ket \propto N^{-\sigma_q}$ where $N$ is the linear size of the system, are related to the level compressibility $\chi$
as $\sigma_q\approx q(1-\chi)[1+q\chi]^{-1}$ for a limited range of $q$; thus providing a way to probe level correlations by means of scattering
experiments.
\end{abstract}
\pacs{05.45.Df, 71.30.+h, 73.23.-b}

\maketitle

\section{Introduction}

Multifractality\cite{MJ98,EM08} appears to be an essential feature of electronic states in disordered systems right at the Anderson--transition which has
been demonstrated in several experiments.\cite{expmuf} The detailed nature of such complex structures received a renewed interest 
and recently many interesting results emerged.\cite{BG11,BG11b,BG11c,ROF11} 

The multifractal dimensions characterizing these states have been obtained mainly using numerical simulations. 
Exact, analytical estimates are available only perturbatively, therefore heuristic relations should help a deeper 
understanding of the complexity of these states. In a recent paper~\cite{EPL12} we have already presented such 
relations especially in case of various random matrix ensembles, hereby we wish to extend these results to different 
observables and to a number of models where either analytical or numerical results are available.
We will demonstrate the robustness of the relations first presented in Ref.~[\onlinecite{EPL12}] giving suggestions for possible experimental
tests, as well.

As mentioned above the spatial fluctuations of the eigenstates are captured in the way how the mean generalized inverse participation numbers
scale with system size
\begin{equation}
          \left\langle \sum_{i=1}^N |\Psi_i|^{2q} \right\rangle \sim N^{-(q-1)D_q} \ ,
\label{Dq}
\end{equation}
where $\left\langle \cdots \right\rangle$ is the average over some states within an
eigenvalue window and over random realizations of the matrix. In the regime of strong localization the scaling is trivial since these
moments are essentially independent of system size resulting in $D_q\to 0$ for all $q$. In the regime of weak disorder, however,
the states appear to be perturbed extended Bloch-states whose moments all scale with the embedding dimension, i.e. $D_q\to d$, for all $q$.
Right at the critical point and even in its vicinity, the $D_q$ dimensions result to be a nonlinear function of the parameter $q$.

Some of these generalized dimensions have an immediate meaning, for instance as $q\to 1$ we arrive at the scaling of the
information entropy of the eigenstates as
\begin{equation}
          \left\langle -\sum_{i=1}^N |\Psi_i|^2 \ln |\Psi_i|^2 \right\rangle \sim D_1 \ln N \ .
\label{D1}
\end{equation}
A further, well--known and widely used dimension is called the correlation dimension $D_2$, which is extracted from the inverse participation number
from Eq.~(\ref{Dq}) using $q=2$. This exponent shows up in the power--law scaling of the density--density correlation function 
in the energy domain as\cite{CK07}
\begin{equation}
\label{correl}
\left\langle |\Psi_{\mu}|^2 |\Psi_{\nu}|^2 \right\rangle \propto 
              \left| \frac{E_0}{E_{\mu}-E_{\nu}} \right|^{1-D_2/d} \ ,
\end{equation}
as well as in the auto--correlation in space as\cite{W80}
\begin{equation}
\left\langle |\Psi_k|^2 |\Psi_l|^2 \right\rangle \propto 
              \left(\frac{N}{|\bm{r}_k-\bm{r}_l|} \right)^{d-D_2} \ ,
\end{equation}
where $\bm{r}_k$ and $\bm{r}_l$ denote the position of sites $k$ and $l$, respectively. In Eq.~(\ref{correl}) the energy scale $E_0$
is of the order of the bandwidth.\cite{CK07}

At the disorder driven Anderson transition not only the eigenstates but also the spectra show unusual behavior. The characterization
of the fluctuations of the spectrum can be done in many ways.\cite{RMT} An often employed quantity, the level 
compressibility $\chi$, in connection to the energy asymptotic limit of the level number variance, is related to medium and long range spectral correlations. Its
definition is given as
\begin{equation} 
          \Sigma^{(2)}(E) = \left\langle n(E)^2 \right\rangle - \left\langle n(E) \right\rangle^2 \sim \chi E \ , 
\end{equation}
using $n(E)$ as the number of states in an interval of length $E$ if $E\gg 1$, i.e. much beyond the mean level spacing. In a metal
the spectral fluctuations can be more-or-less described by standard random matrix theory and therefore the compressibility 
vanishes, $\chi\to 0$, while in the extreme of strong disorder, i.e. when the spectrum is uncorrelated, Poisson statistics
yields $\chi=1$ giving $\chi$ a character of order parameter. Right at the transition multifractality shows up in statistical fluctuations
that are intermediate between these two extremes, i.e. $0<\chi <1$. 

An even more interesting result is that the quantities which describe the statistical fluctuations of the spectra and those for the eigenstates may be related as first pointed out  long ago
in Ref.~[\onlinecite{CKL96}]. For the critical three--dimensional Anderson transition and the two--dimensional quantum--Hall transition 
it has been shown earlier that 
\begin{equation}
          2\chi + D_2/d = 1 \ .
\label{chiD2}
\end{equation}
This relation should obviously hold approximately only since the range of the correlation dimension and that of 
the level compressibility are limited as $0\leq D_2/d\leq 1$ and $0\leq\chi\leq 1$, respectively, leaving the 
validity of (\ref{chiD2}) to the limit of weak--multifractality; i.e. if $D_2\to d$ then $\chi\to 0$.

More recently, Bogomolny and Giraud\cite{BG11} showed that for systems at criticality in $d$--dimensions 
$D_1$ and $\chi$ can be related in a very simple way:
\begin{equation}
              \chi + D_1/d = 1 \ .
\label{chiD1}
\end{equation}
Furthermore, using as evidence various critical random matrix ensembles, in Ref.~[\onlinecite{BG11}] it was 
shown that
\begin{equation}
          \frac{D_q}{d} = \left\{
                  \begin{array}{ll}
                             \displaystyle
                       \frac{\Gamma(q-1/2)}{\sqrt{\pi}\Gamma(q)}(1-\chi) \ , &
                        1-\chi \ll 1 \\
                        1-q\chi \ , & \chi \ll 1
                  \end{array}
\right.  \ .
\label{Dqchith}
\end{equation}
The latter relation seemed to be valid for any type of multifractality.

 Looking at Eqs.~(\ref{chiD2}) and (\ref{chiD1}) it seems evident that there may exist further relations between generalized dimensions and the level
compressibility, moreover, one may conjecture that the generalized dimensions of the multifractal spectrum must be also intimately linked together.

In Ref.~[\onlinecite{EPL12}] we have presented evidences that indeed a series of relations between various generalized dimensions, $D_q$ and 
$D_{q'}$, and the level compressibility $\chi$ exist allowing for a generalization that contains Eq.~(\ref{chiD1}) exactly and Eq.~(\ref{chiD2}) 
in the appropriate limit. In order to prove that, numerical simulations of various critical random matrix ensembles have been used. 
In the present work we wish to give a series of details that were not included in our earlier work. Moreover, by the use of our relations 
between generalized dimensions we state a clear link between the spectral and scattering properties of disordered systems at the metal-insulator transition.
Finally, by exploring both, a deterministic model having a self-similar potential that produces multifractal eigenstates and multifractal objects expressly 
constructed by the use of iteration algorithms, we also show that our results are not restricted to random matrix models. 

\section{Model and heuristic relations}

In Ref.~[\onlinecite{BG11}] Eqs.~(\ref{chiD1}) and (\ref{Dqchith}) were shown to be correct numerically for the Power-Law Banded Random Matrix 
(PBRM) model\cite{EM08,MFDQS96,Mirlin00} at criticality. Below we will make use of this model to briefly present the heuristic relations already
published in our earlier work. Furthermore, we expand the applicability of our approach to an extended range of multifractal dimensions
as well as to scattering quantities.

The PBRM model describes one--dimensional (1d) samples of length $N$ with random long-range hoppings. This model is represented by
$N\times N$ real symmetric ($\beta=1$) or complex hermitian ($\beta=2$) matrices whose elements are statistically independent random variables
drawn from a normal distribution with zero mean and a variance given by $\langle  |H_{mm}|^2 \rangle =\beta^{-1}$ and
\begin{equation}
          \langle  |H_{mn}|^2 \rangle = \frac{1}{2} \frac{1}{1+\left[\sin\left( \pi|m-n|/N \right)/(\pi b/N) \right]^{2\mu}} \ ,
\label{PBRMp}
\end{equation}
where $b$ and $\mu$ are parameters. In Eq.~(\ref{PBRMp}) the PBRM model is in its periodic version; i.e. the 1d sample is in a ring geometry.
Theoretical considerations\cite{EM08,MFDQS96,Mirlin00,KT00} and detailed numerical investigations\cite{EM08,EM00b,V03} have verified that the PBRM
model undergoes a transition at $\mu=1$ from localized states for $\mu >1$ to delocalized states for $\mu < 1$. This transition 
shows key features of the disorder driven Anderson metal-insulator transition,\cite{EM08} including multifractality of eigenfunctions and 
non-trivial spectral statistics. Thus the PBRM model possesses a line of critical points $b\in (0,\infty)$ in the case of $\mu=1$. In the 
following we will focus on the PBRM model at criticality, $\mu=1$. By tuning the parameter $b$, from $b\ll 1$ to $b\gg 1$, the states 
cross over from the nature of strong--multifractality ($D_q \to 0$) which corresponds to localized--like or insulator--like states to 
weak--multifractality ($D_q \to 1$) showing rather extended, i.e. metallic--like states. Meanwhile at the true Anderson transition 
in $d=3$ or at the integer quantum--Hall transition in $d=2$, the states belong to the weakly multifractal regime, i.e.
$d-D_2\ll d$, the PBRM model allows for an investigation without such a limitation. The evolution of the generalized dimensions as a 
function of the parameter $b$ therefore represent this behavior, i.e. $D_q\to 1$ for $b\gg 1$ and in the limit of $b\ll 1$ the
multifractal dimensions vanish as $D_q\sim b$.\cite{EM08,Mirlin00}

\begin{figure}[t]
\centerline{\includegraphics[width=7cm]{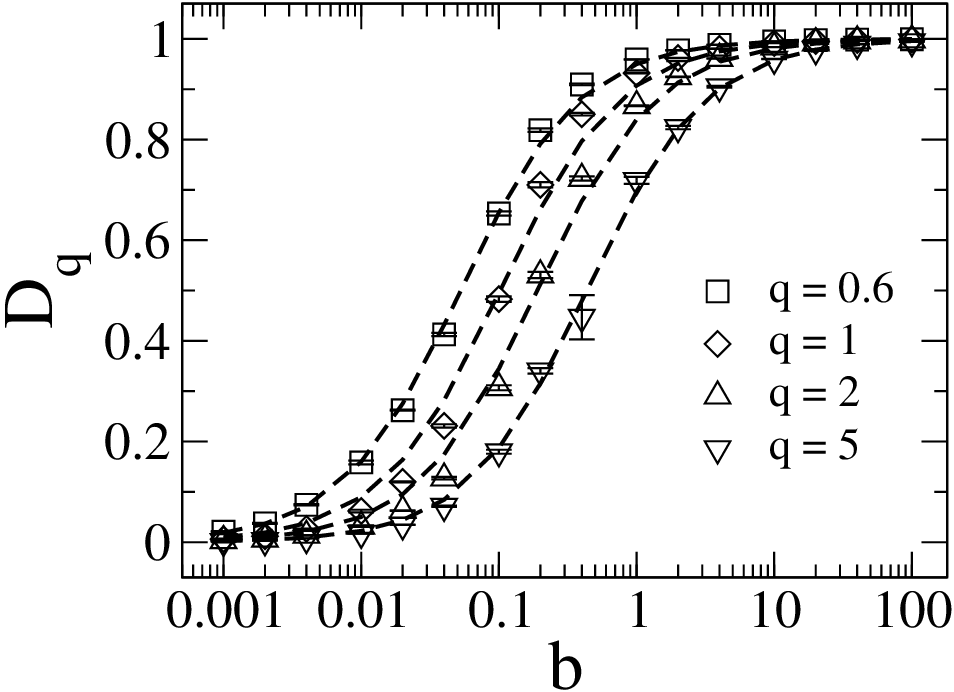}}
\caption{$D_q$ as a function of $b$ for the PBRM model at criticality with $\beta=2$. The dashed lines are fits of the numerical data with
Eq.~(\ref{Dqofb}).}
\label{Fig1}
\end{figure}

\begin{figure}[t]
\centerline{\includegraphics[width=7cm]{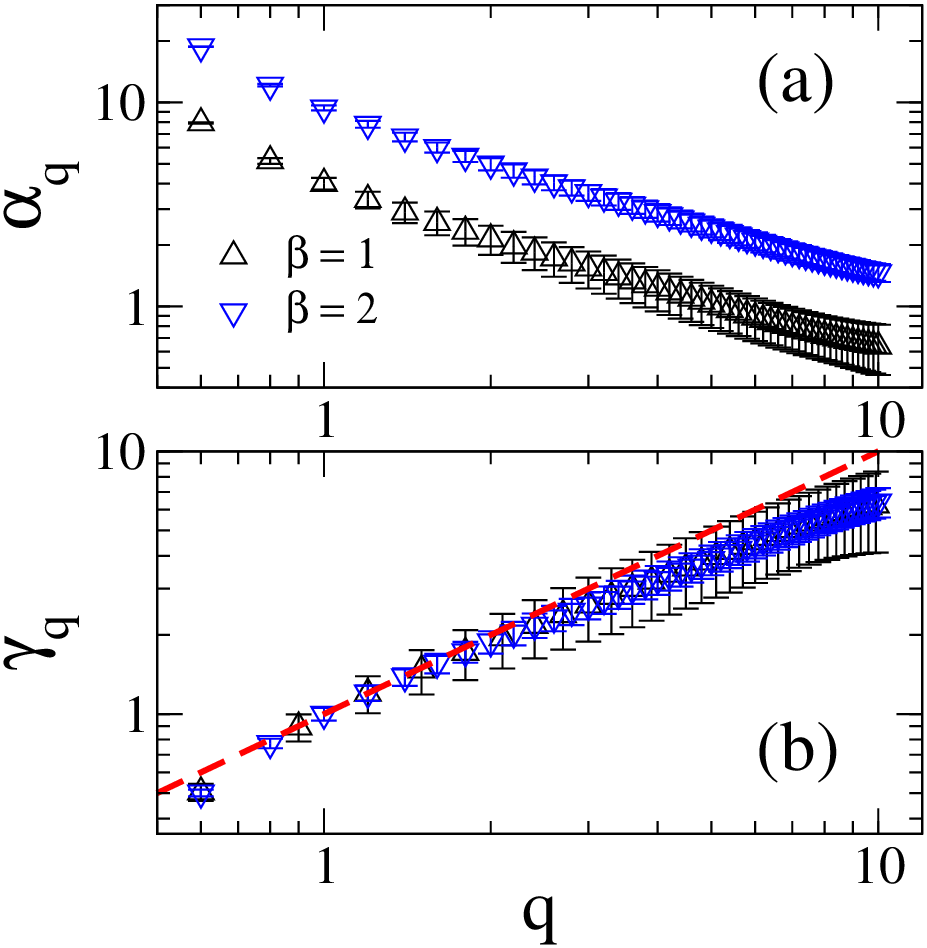}}
\caption{(Color online) (a) $\alpha_q$ and (b) $\gamma_q = \alpha_1/\alpha_q$ as a function of $q$ for the PBRM model at criticality with $\beta=1$ and $\beta=2$. 
The red dashed line in (b) equal to $q$ is plotted to guide the eye. The error bars in (a) are the rms error of the fittings.}
\label{Fig2}
\end{figure}

The multifractal dimensions, especially $D_1$ and $D_2$ and their dependence on the parameter $b$ in the case of the PBRM model, have
been at the focus of several of our works.\cite{MTC06,MV06} In those works it has been demonstrated that simple phenomenological
relations can be identified: $D_1 \approx [1+(\alpha_1 b)^{-1}]^{-1}$ and $D_2 \approx [1+(\alpha_2 b)^{-1}]^{-1}$ where $\alpha_{1,2}$ are 
fitting constants. These continuous functions are trivial interpolations between the limiting cases of low--$b$ and large--$b$, 
taking the half of the harmonic mean of the two cases. In Ref.~[\onlinecite{EPL12}] we generalized and propose the following 
heuristic expression for an extended range of the parameter $q$
\begin{equation}
          D_q \approx \left[ 1+(\alpha_q b)^{-1} \right]^{-1} \ ,
\label{Dqofb}
\end{equation}
as a global fit for the multifractal dimensions $D_q$ of the PBRM model for both symmetry classes, $\beta=1$ and $\beta=2$. 
In Fig.~\ref{Fig1} we show fits of Eq.~(\ref{Dqofb}) to numerically obtained $D_q$'s as a function of $b$ for some values of $q$
for the PBRM model with $\beta=2$ (the case $\beta=1$ has been already reported in Ref.~[\onlinecite{EPL12}]) and in Fig.~\ref{Fig2}(a) we 
plot the values of $\alpha_q$ extracted from the fittings; here, for comparison purposes, we report both cases: $\beta=1$ and $\beta=2$. 
We observe that Eq.~(\ref{Dqofb}) fits reasonably well the numerical $D_q$ for $q>1/2$. It is important to stress that
Eq.~(\ref{Dqofb}) reproduces well the $b$-dependencies predicted analytically\cite{EM08} for the limits $b\ll 1$ and $b\gg 1$.

The multifractal dimensions of Fig.~\ref{Fig1} were extracted from the linear fit of the logarithm of the inverse mean eigenfunction participation numbers
versus the logarithm of $N$, see Eq.~(\ref{Dq}). $D_1$ was extracted from the linear fit of the mean eigenfunction entropy versus the logarithm 
of $N$, see Eq.~(\ref{D1}). We used system sizes of $N=2^n$ with $8\le n\le 13$. The average was performed over $2^{n-3}$ eigenvectors
with eigenvalues around the band center of $2^{16-n}$ realizations of the random matrices.

We have also verified that
\begin{equation}
          \chi \approx \left( 1+\alpha_\chi b \right)^{-1}  \ ,
\label{chiofb}
\end{equation}
reproduces qualitatively well the $b$-dependencies predicted analytically\cite{EM08,KYC} in the small-- and large--$b$ limits:
\begin{equation}
    \chi = \left\{ 
    \begin{array}{ll}
              \left.
              \begin{array}{ll}
                             1-4b \ ,                                                                      & \beta = 1 \\
                      1-\pi\sqrt{2}b + \frac{4}{3}\left(2-\sqrt{3}\right)\pi^2b^2 \ , & \beta = 2
              \end{array}
              \right.
              \quad & b \ll 1 \\
              \displaystyle \frac{1}{2\beta \pi b} & b \gg 1
    \end{array}
             \right. \ .
\label{chithPBRM}
\end{equation}
Moreover, the parameter $\alpha_\chi$ must be equal to $\alpha_1$ for Eqs.~(\ref{Dqofb}) and (\ref{chiofb}) 
to fulfill relation (\ref{chiD1}).

As a consequence of equating $b$ in Eqs.~(\ref{Dqofb}) and (\ref{chiofb}) we get a direct relation
\begin{equation}
          \chi \approx (1-D_q) \left[ 1+(\gamma_q-1)D_q \right]^{-1} \ ,
\label{chiofgamma}
\end{equation}
with $\gamma_q = \alpha_1/\alpha_q$. We observed that $\gamma_q \approx q$ in the range $0.8<q<2.5$, 
as plotted in Fig.~\ref{Fig2}(b), so in this range of $q$ values we can write simplified relations between $\chi$ and $D_q$:
\begin{equation}
         \chi \approx \frac{1-D_q}{1+(q-1)D_q} \quad \mbox{and} \quad D_q \approx \frac{1-\chi}{1+(q-1)\chi} \ .
\label{Dqofchi}
\end{equation}
The expression for $D_q$ in Eq.~(\ref{Dqofchi}) reproduces Eq.~(\ref{Dqchith}) exactly for $q=1$ and $q=2$ and approximately for $1<q<2.5$.
Moreover, Eq.~(\ref{Dqofchi}) combined with Eq.~(\ref{chiD1}) allows us to express any $D_q$ in terms of, for example, $D_1$:
\begin{equation}
          D_q \approx D_1 \left[ q+(1-q)D_1 \right]^{-1} \ .
\label{DqofD1}
\end{equation}

We also noticed that by equating $\chi$ for different $D_q$'s from Eq.~(\ref{Dqofchi}) we could get recursive relations among them:
\begin{equation}
     q'D_{q'} (1-D_{q'})^{-1} = qD_q(1-D_q)^{-1} \ ,
\label{DqpDq}
\end{equation}
which in case of taking $q'=q+1$ leads to 
\begin{equation}
     D_{q+1} = qD_q(1+q-D_q)^{-1} \ .
\label{DqpofDq}
\end{equation}
Notice that all these relations can be expressed using the fact that the ratio $qD_q/(1-D_q)$ is independent of $q$.
These expressions also provide a relation between the correlation dimension and the information dimension or between the
correlation dimension and the compressibility of the spectrum:
\begin{equation}
          D_2 = D_1\left( 2-D_1 \right)^{-1} = \left( 1-\chi \right) \left( 1+\chi \right)^{-1} \ .
\label{D2ofD1}
\end{equation}
It is relevant to add that in the weak multifractal regime, i.e. when $\chi\to 1$, Eq.~(\ref{D2ofD1}) reproduces the relation given in Eq.~(\ref{chiD2}) 
with $d=1$, reported in [\onlinecite{CKL96}].

\subsection{The case $q<1/2$}

For $q<1/2$, Eq.~(\ref{Dqofb}) cannot be directly applied. However, the regime $q<1/2$ could also be explored within our
approach by the combination of Eq.~(\ref{Dqofb}) and  the symmetry relation\cite{MFME06}
\begin{equation} 
      \Delta_q = \Delta_{1-q} \quad \mbox{with} \quad  \Delta_q = D_q(q-1)-d(q-1) \ ,
\label{sym}
\end{equation} 
implying that it is possible to link the multifractal dimensions with indexes $q<1/2$ to those with $q>1/2$. Therefore, we get 
\begin{equation}
          D_q \approx \frac{1-2q}{1-q} + \frac{q}{1-q}\left( \frac{\alpha_{1-q}b}{1+\alpha_{1-q}b} \right) \ ,
\label{D1_qofb}
\end{equation}
for $q<1/2$; that is, once we know the coefficients $\alpha_q$ for $q>1/2$ we can use them to get $D_q$ for $q<1/2$. Moreover, by the use of 
Eq.~(\ref{DqpofDq}) it is possible to write $D_q$, for $q<1/2$, as a function of any specific $D_q$ with $q>1/2$. For example, 
\begin{equation}
       D_q \approx \frac{1-2q}{1-q} + \frac{q}{1-q}\left( \frac{D_1}{1+q(D_1-1)} \right) \ ,
\label{D1_qofD1}
\end{equation}
provides $D_q$ for $q<1/2$ in terms of the information dimension. Moreover, we can write down relations between $\chi$ and $D_q$ with $q<1/2$: 
\begin{equation}
        \chi \approx \frac{1-D_q}{q(2-D_q)} \quad \quad \mbox{and} \quad \quad D_q \approx \frac{1-2q\chi}{1-q\chi} \ .
\label{Dqofchi_1}
\label{Dq_1ofchi}
\end{equation}

Finally, it is important to stress that Eqs.~(\ref{D1_qofb}-\ref{D1_qofD1}) get the form
\begin{equation}
      D_q = (1-2q)(1-q)^{-1}
\label{D1_qb0}
\end{equation}
in the limit $b\to 0$, which has been derived analytically in Refs.~[\onlinecite{MFME06,MG10,VK,BG11c}].

\subsection{Wigner delay times}

As mentioned in the Introduction, the modeling and analysis of multifractal states in disordered systems at the Anderson--transition
has been a subject of intensive research activity for many decades.\cite{MJ98,EM08} Moreover, since the properties of the closed system, i.e. the fractality
of the eigenstates, strongly influence the scattering and transport properties of the corresponding open system, the interest has also
been extended to critical {\it scattering} systems. In particular, much attention has been focused on the probability
distribution functions of the resonance widths and Wigner delay times,~\cite{MV06,MK05,OF05,MFME06,OKG03,F03,TC99,KW02,KW02b,WMK06}
as well as the transmission or dimensionless
conductance.~\cite{S90,M94,M99,SO97,SOK00,SMO00,WLS98,RS01,RMS001,TM02,SM05,JMZ99,SZ02,MG09a,MG09b,MGV10,MMV10}

Among many relevant results, here we want to focus on those related to Wigner delay times $\tau_{\tbox W}$ and recall that 
\begin{itemize}
   \item[(i)] for disordered systems at criticality the inverse moments of Wigner delay times $\bra \tau_{\tbox W}^{-q} \ket$ scale as\cite{MFME06,MK05,OF05} 
\begin{equation}
   \bra \tau_{\tbox W}^{-q} \ket \propto N^{-\sigma_q} \quad \mbox{where} \quad \sigma_q \equiv qD_{q+1} \ ;
\label{sigmaqofDq}
\end{equation}
   \item[(ii)] for the PBRM model at criticality the typical values of the Wigner delay times scale as\cite{MV06} 
\begin{equation}
   \tau_{\tbox W}^{\tbox{typ}} \propto N^{\sigma_\tau} \quad \mbox{where} \quad \sigma_\tau = D_1 \ ,
\label{sigmaofD1}
\end{equation} 
with $\tau_{\tbox W}^{\tbox{typ}}\equiv \exp\bra\ln\tau_{\tbox W}\ket$.
\end{itemize}

Note that Eqs.~(\ref{sigmaqofDq}) and (\ref{sigmaofD1}) provide a way to probe the properties of a critical system (i.e. the fractality
of its eigenstates) by means of scattering experiments. Moreover, we can also relate spectral properties to scattering properties by 
\begin{itemize}
   \item[(i)] combining Eqs.~(\ref{Dqofchi}) and (\ref{sigmaqofDq}):
   \begin{equation}
          \sigma_q \approx \frac{q(1-\chi)}{1+q\chi} \quad \mbox{where} \quad\ \chi \approx \frac{q - \sigma_q}{q(\sigma_q+1)} \ ;
   \label{sigmaqofchi}
   \end{equation}
   \item[(ii)] combining Eqs.~(\ref{D2ofD1}) and (\ref{sigmaofD1}):
   \begin{equation}
           \sigma_\tau = 1-\chi \ .
   \label{sigmaofchi}
   \end{equation} 
\end{itemize}

Also, we can express any $\sigma_q$ as a function of, say, $\sigma_1$:
\begin{equation}
   \sigma_q = 2q\sigma_1[(1+q)+(1-q)\sigma_1]^{-1} \ .
\label{sqofs1}
\end{equation} 
Finally, note that we can obtain recursive relations for $\sigma_q$'s in analogy to Eq.~(\ref{DqpDq}):  
\begin{equation}
    (q'+1)\sigma_{q'}(q'-\sigma_{q'})^{-1} = (q+1)\sigma_q(q-\sigma_q)^{-1} \ ,
\label{sqpsq}
\end{equation} 
or
\begin{equation}
   \sigma_{q'} = q'(q+1)\sigma_q[q(q'+1)+(q-q')\sigma_q]^{-1} \ ,
\label{sqpofsq}
\end{equation} 
which leads to
\begin{equation}
   \sigma_{q+1} = (q+1)^2\sigma_q[q(q+2)-\sigma_q]^{-1} \ ,
\label{sqm1ofsq}
\end{equation} 
when $q'=q+1$.

\begin{figure}[t]
\centerline{\includegraphics[width=7cm]{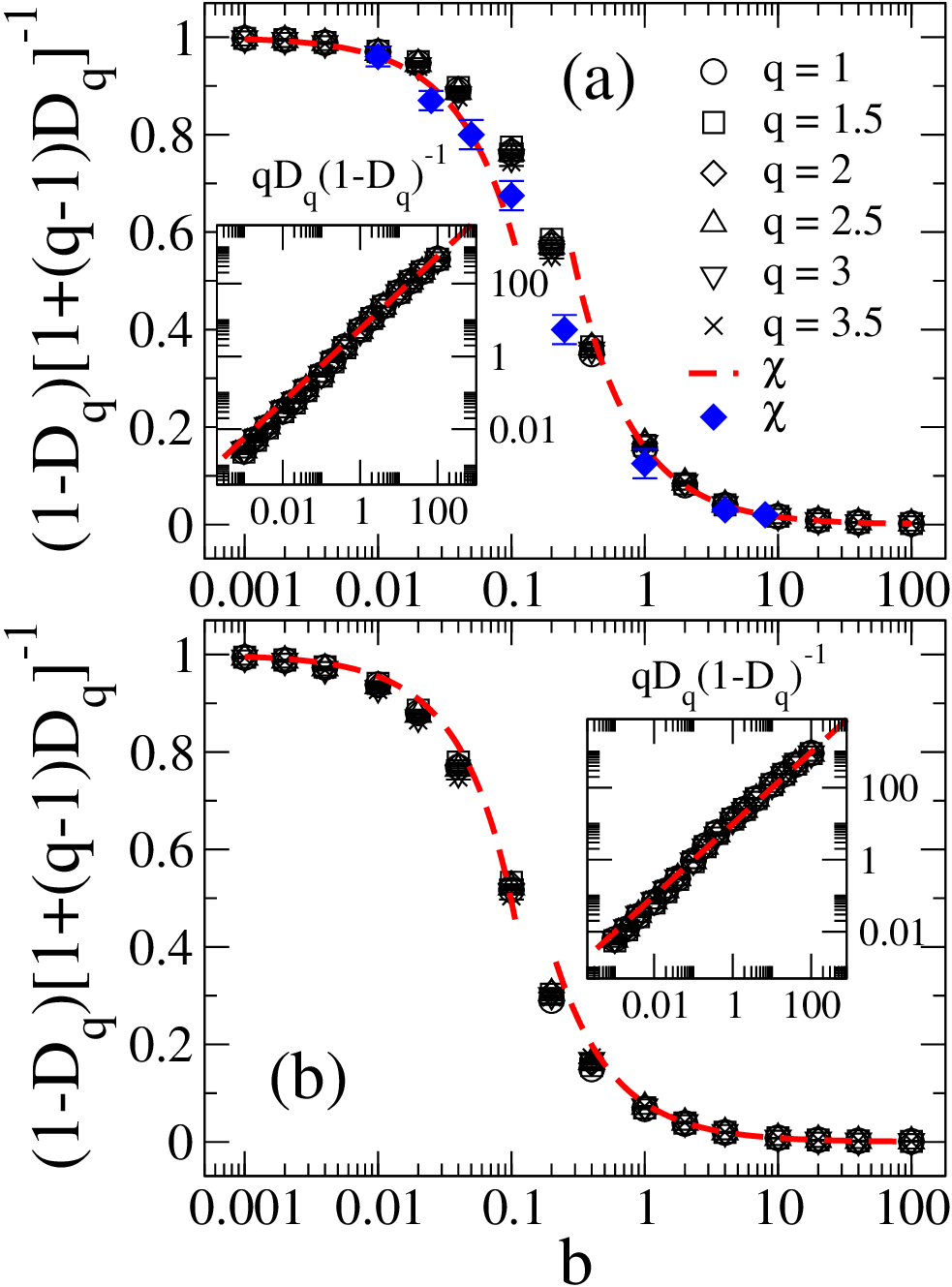}}
\caption{(Color online) $(1-D_q)[1+(q-1)D_q]^{-1}$ [see Eq.~(\ref{Dqofchi})] as a function of $b$ for the PBRM model at criticality with (a) $\beta=1$ and (b) $\beta=2$.
The red dashed lines are the analytical predictions for $\chi$ given by Eq.~(\ref{chithPBRM}). The blue symbols in (a) are independent numerically obtained values of 
$\chi$ (taken from Ref.~[\onlinecite{KS}]). Insets: $qD_q(1-D_q)^{-1}$ as a function of $b$; see Eq.~(\ref{DqpDqPBRM}). The red dashed line equal to $\alpha_1b$ 
is plotted to guide the eye. (a) $\alpha_1 = 4$ and (b) $\alpha_1 = 9.4$ were used.}
\label{Fig3}
\end{figure}
\begin{figure}[t]
\centerline{\includegraphics[width=7cm]{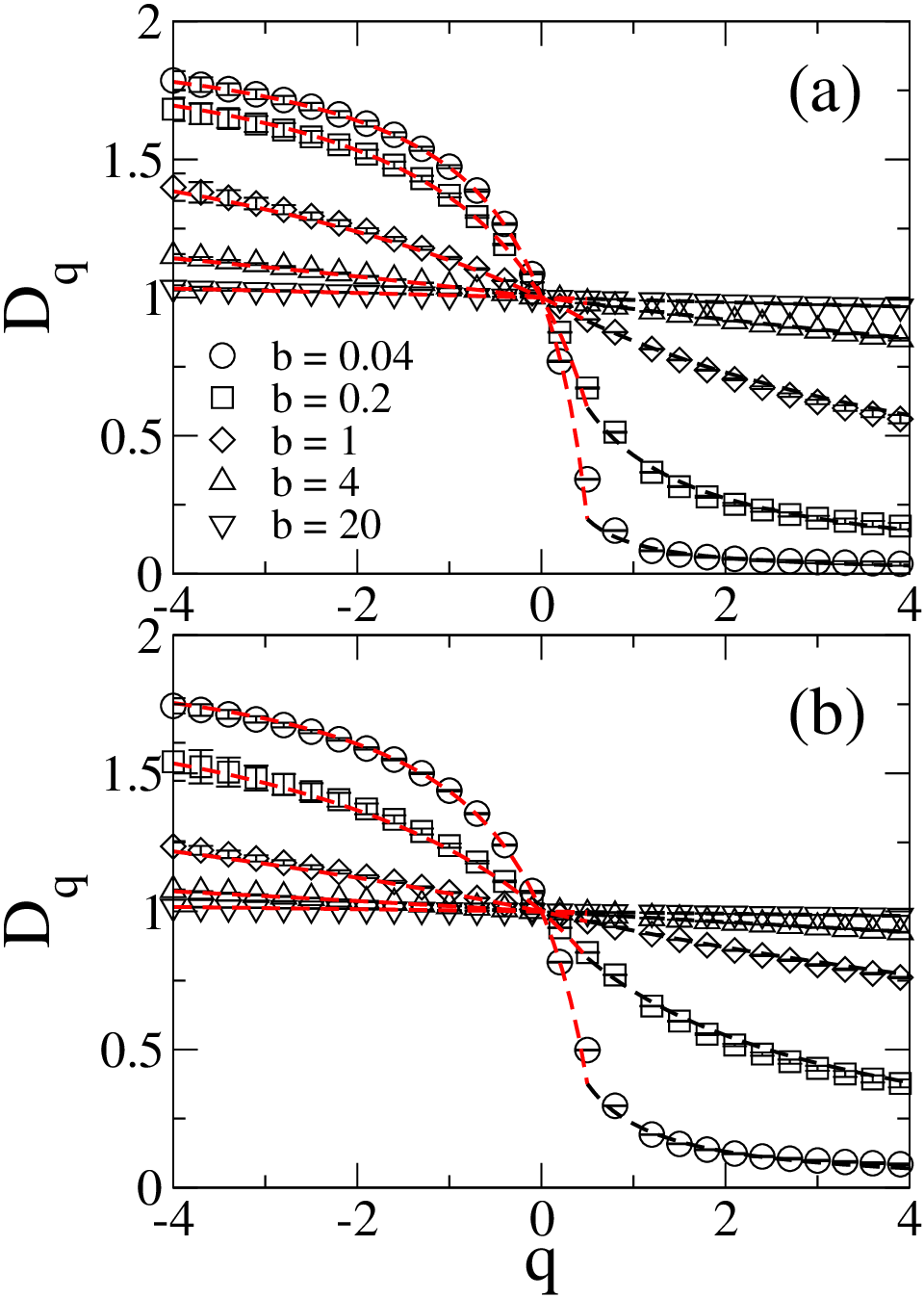}}
\caption{$D_q$ as a function of $q$ for the PBRM model at criticality with (a) $\beta=1$ and (b) $\beta=2$. Black [Red] 
dashed lines are Eq.~(\ref{DqofD1}) [Eq.~(\ref{D1_qofD1})]. We used (a) [(b)] $D_1=0.109$, 0.429, 0.843, 0.958, and 0.991
[$D_1=0.230$, 0.711, 0.931, 0.980, and 0.996].}
\label{Fig4}
\end{figure}

\section{Numerical results for the PBRM model}

In this section we present numerical justification of our analytical relations derived above using the PBRM model. 

\subsection{Multifractal exponents}

In Fig.~\ref{Fig3} we plot $(1-D_q)[1+(q-1)D_q]^{-1}$ as a function of $b$ for several values of $q$ for the PBRM model at criticality 
with $\beta=1$ and $\beta=2$ and observe good correspondence with the analytical prediction for $\chi$; that is, we verify the validity 
of Eq.~(\ref{Dqofchi}). For completeness in Fig.~\ref{Fig3}(a) we also include independent numerically obtained values of $\chi$ (taken from 
Ref.~[\onlinecite{KS}]). In the insets of Fig.~\ref{Fig3} we plot $qD_q(1-D_q)^{-1}$ as a function of $b$, see Eq.~(\ref{DqpDq}), which 
for the PBRM model acquires the simple form
\begin{equation}
    qD_q\left( 1-D_q \right)^{-1} \approx \alpha_1b \ .
\label{DqpDqPBRM}
\end{equation}
The fact that all curves $qD_q\left( 1-D_q \right)^{-1}$ vs $b$ fall one on top of the other makes evident its independence of $q$.

In Fig.~\ref{Fig4} we present $D_q$ as a function of $q$ for the PBRM model with $\beta=1$ and $\beta=2$ for some values of $b$.
As black and red dashed lines we also include Eqs.~(\ref{DqofD1}), for $q>1/2$, and (\ref{D1_qofD1}), for $q<1/2$, respectively. 
In both equations we used values of $D_1$ we have obtained numerically. We observe very good correspondence between the numerical data and
Eqs.~(\ref{D1_qofD1}) and (\ref{DqofD1}) mainly for $-4<q<0.2$ and $0.8<q<4$, respectively.

\subsection{Wigner delay times}

We obtain Wigner delay times $\tau_{\tbox W}$ by turning the isolated system, represented by the PBRM model, into a scattering one by attaching 
one semi-infinite single-channel lead using perfect coupling. Since we are dealing here with the periodic version of the PBRM model, all sites 
are bulk sites and the place at which we attach the lead is irrelevant. To compute $\tau_{\tbox W}$ we use the 
{\it effective Hamiltonian approach} described in Refs.~[\onlinecite{MK05,MV06}]. For statistical processing a large number 
of disorder realizations is used. Each disorder realization gives one value of $\tau_{\tbox W}$. We used $N=50$, 100, 200, 400, and 800 
getting $10^6$, $10^6$, $10^5$, $10^5$, and $10^4$ values of $\tau_{\tbox W}$, respectively. Then, the exponents $\sigma_q$ [$\sigma_\tau$] were extracted 
from the linear fit of the logarithm of the averaged inverse moments of Wigner delay times [typical Wigner delay times]
{\it vs.} the logarithm of $N$, see Eq.~(\ref{sigmaqofDq}) [Eq.~(\ref{sigmaofD1})]. We concentrate here on the PBRM model with $\beta=1$ only.
\begin{figure}[t]
\centerline{\includegraphics[width=7cm]{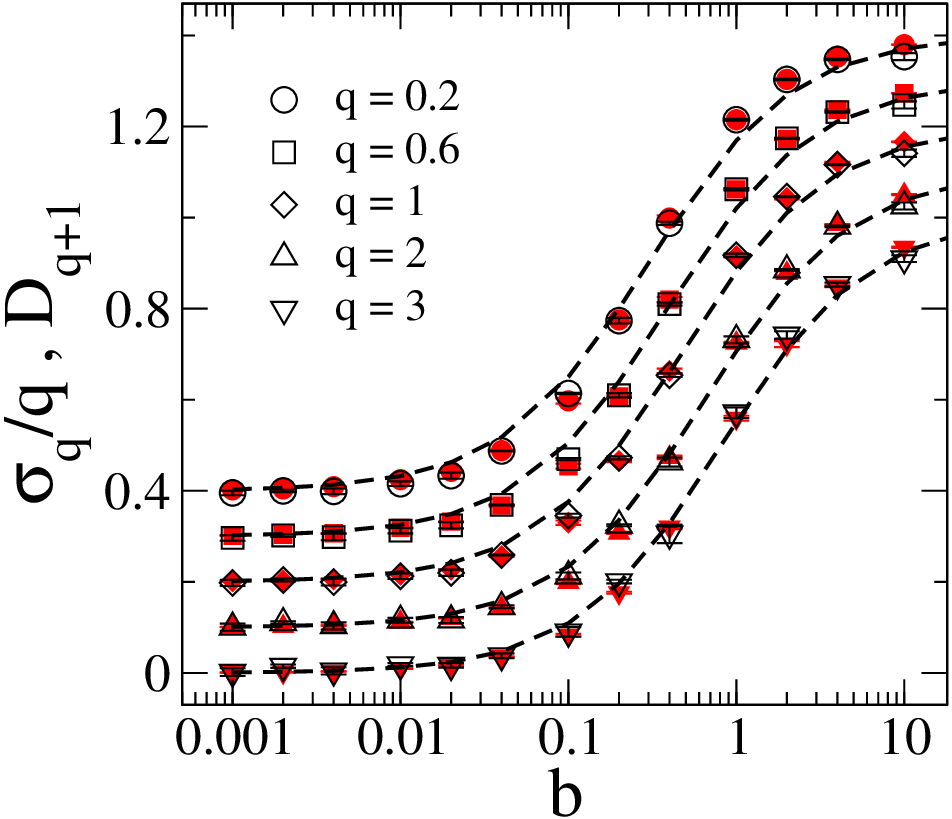}}
\caption{(Color online) $\sigma_q/q$ (black open symbols) and $D_{q+1}$ (red full symbols) as a function of $b$ for the PBRM model at criticality 
with $\beta=1$. The dashed lines are Eq.~(\ref{sigmaqofb}); we used the values of $\alpha_{q+1}$ reported in Fig.~\ref{Fig2}(a).
For clarity the data for $q<3$ was displaced upwards.}
\label{Fig5}
\end{figure}

We start by noticing that if we combine Eqs.~(\ref{Dqofb}) and (\ref{sigmaqofDq}) we get a heuristic expression for $\sigma_q$
as a function of $b$:
\begin{equation}
    \sigma_q \approx \frac{q}{1+(\alpha_{q+1}b)^{-1}} \ .
\label{sigmaqofb}
\end{equation}
In Fig.~\ref{Fig5} we compare Eq.~(\ref{sigmaqofb}) to the numerically obtained $\sigma_q$ as a function of $b$ for some values of $q$. 
We observe that Eq.~(\ref{sigmaqofb}) fits reasonably well the numerical $\sigma_q$ for $q\ge 0.1$. 
In Fig.~\ref{Fig5} we also include independent numerically obtained values of $D_q$ that further verifies the validity of relation 
(\ref{sigmaqofDq}) [\onlinecite{MFME06,MK05,OF05}].

\begin{figure}[t]
\centerline{\includegraphics[width=7cm]{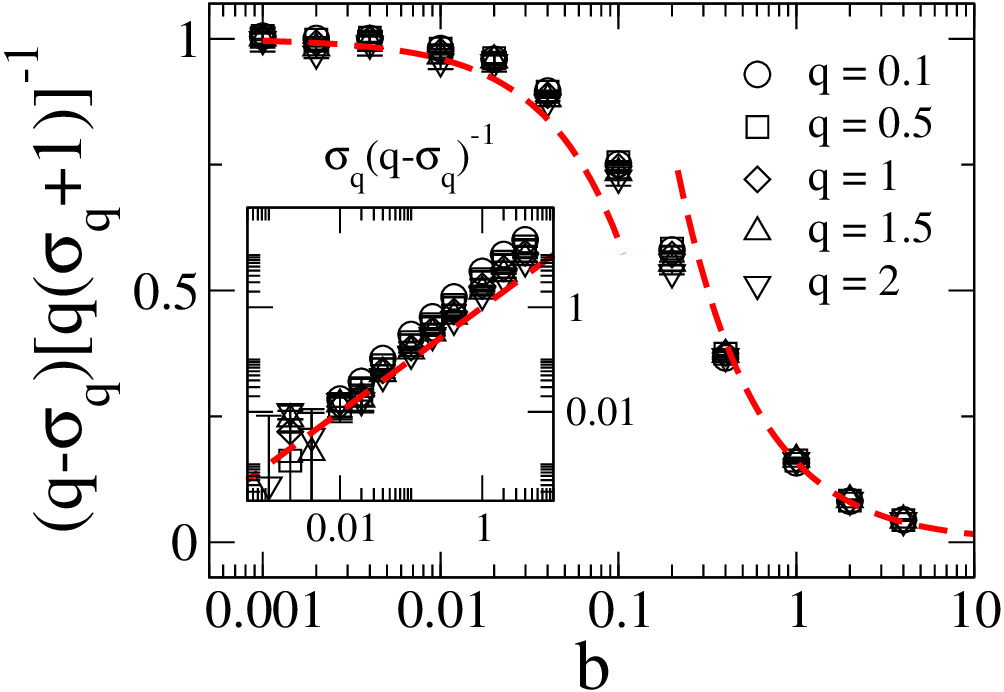}}
\caption{(Color online) $(q-\sigma_q)[q(\sigma_q+1)]^{-1}$ [see Eq.~(\ref{sigmaqofchi})] as a function of $b$ for the PBRM model at criticality with $\beta=1$.
The red dashed lines are the analytical prediction for $\chi$ given by Eq.~(\ref{chithPBRM}). Inset: $\sigma_q(q-\sigma_q)^{-1}$ as a function 
of $b$; see Eq.~(\ref{sigmaqofalpha}). The red dashed line equal to $b$ is plotted to guide the eye.}
\label{Fig6}
\end{figure}

In Fig.~\ref{Fig6} we plot $(q-\sigma_q)[q(\sigma_q+1)]^{-1}$ as a function of $b$ for the PBRM model at criticality; 
that is, we verify the validity of Eq.~(\ref{sigmaqofchi}). We also plot the analytical prediction for $\chi$ given in 
Eq.~(\ref{chithPBRM}) and observe good correspondence with the numerical data. In the inset we plot $\sigma_q/(q-\sigma_q)$ as a 
function of $b$, see Eq.~(\ref{sqpsq}), which for the PBRM model acquires the simple form 
\begin{equation}
     \sigma_q(q-\sigma_q)^{-1} \approx \alpha_{q+1}b \ .
\label{sigmaqofalpha}
\end{equation}

\begin{figure}[t]
\centerline{\includegraphics[width=6cm]{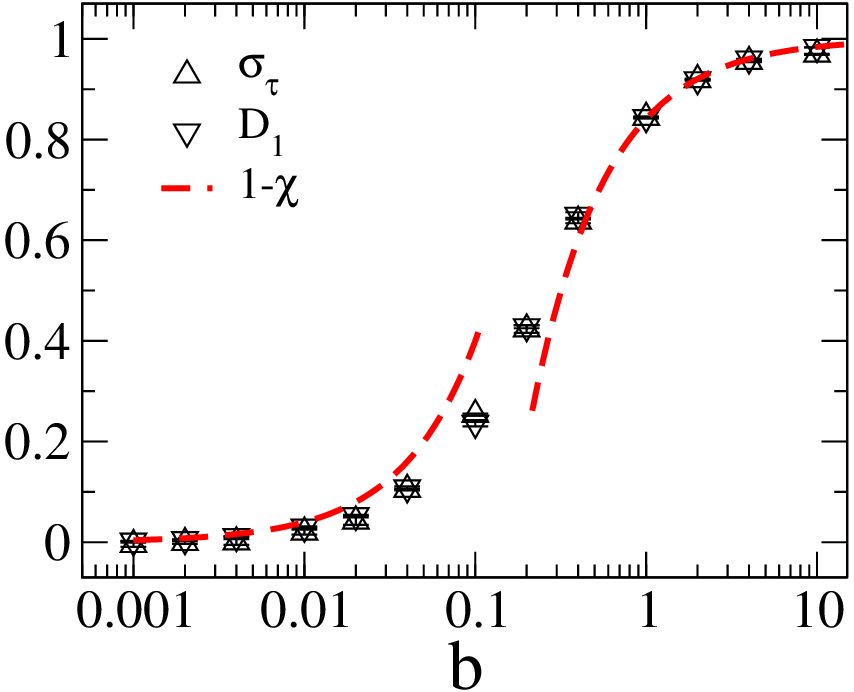}}
\caption{(Color online) $\sigma_\tau$ as a function of $b$ for the PBRM model at criticality with $\beta=1$. We include the numerically 
obtained $D_1$, see Eq.~(\ref{sigmaofD1}). The red dashed lines are $1-\chi$, see Eq.~(\ref{sigmaofchi}).}
\label{Fig7}
\end{figure}

Finally, in Fig.~\ref{Fig7} we show $\sigma_\tau$ as a function of $b$ for the PBRM model at criticality with $\beta=1$.
To test the validity of Eq.~(\ref{sigmaofchi}) we compare $\sigma_\tau$ with the numerically obtained 
$D_1$ and with the theoretical prediction for $1-\chi$. We again observe good correspondence.

\section{Other critical ensembles}

Remember that relations (\ref{Dqofchi}-\ref{D2ofD1}) were obtained form the combination of Eqs.~(\ref{Dqofb}) and (\ref{chiofb}). 
That is, relations (\ref{Dqofchi}-\ref{D2ofD1}) are expected to work in particular for the PBRM model at criticality. However, 
Eqs.~(\ref{Dqofchi}) reproduce Eqs.~(\ref{chiD1}) and (\ref{Dqchith}), which were shown to be valid for the PBRM model but also for other 
critical ensembles.\cite{BG11} Therefore the question arises to which extent relations (\ref{Dqofchi}-\ref{D2ofD1}) are valid for critical 
ensembles different to the PBRM model. So, in the following we verify the validity of Eqs.~(\ref{Dqofchi}-\ref{D2ofD1}) for other well known 
critical ensembles.

\subsection{Calogero-Moser ensembles}

The Calogero--Moser (CM) $N$-particle systems yield three ensembles of $N\times N$ Hermitian matrices of the form\cite{BG11b,BG11c}
\begin{equation}
    H_{mn} = p_m\delta_{mn} + i g(1-\delta_{mn}) V(m-n) \ ,
\label{CME}
\end{equation}
where $p_m$ are independent Gaussian random variables with zero mean and unit variance, $g$ is a free parameter which drives the
multifractality of the eigenstates, and $V(m-n)$ is one of the three following functions
\[
    \frac{1}{m-n} \ , \quad \frac{1}{N\sinh[(m-n)/N]} \ , \quad \frac{1}{N\sin[(m-n)/N]} \ .
\]
These ensembles were denoted as\cite{BG11b,BG11c} CMR, CMH, and CMT, respectively.

\begin{figure}[t]
\centerline{\includegraphics[width=7cm]{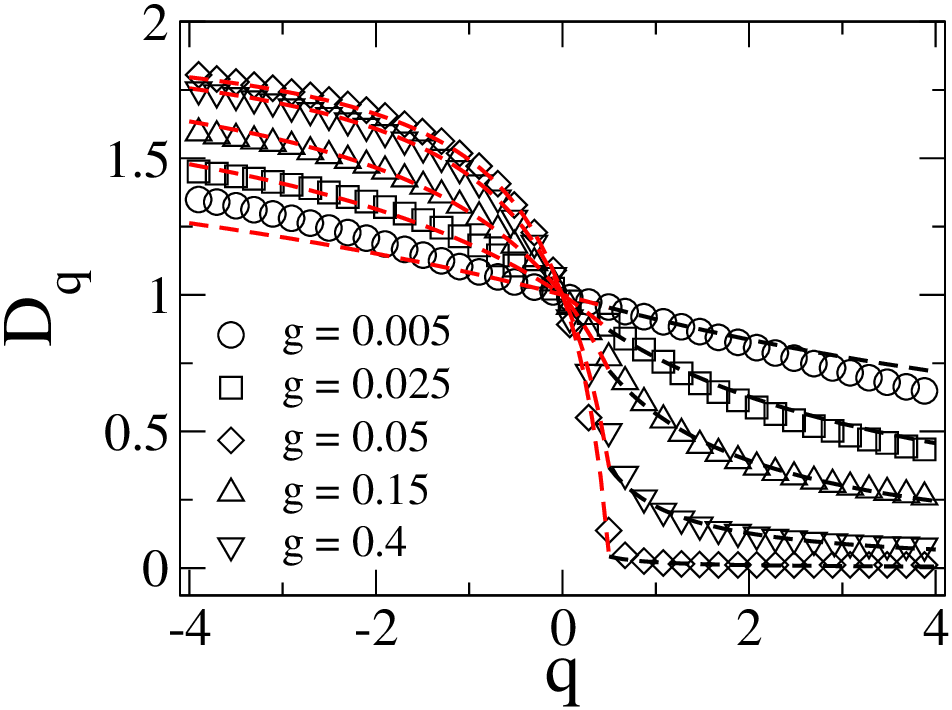}}
\caption{(Color online) $D_q$ as a function of $q$ for the CMR ensemble for several values of $g$.
The data was taken from Ref.~[\onlinecite{BG11c}]. Black [Red] dashed lines are Eq.~(\ref{DqofD1}) [Eq.~(\ref{D1_qofD1})].
We used $D_1=0.911$, 0.771, 0.565, 0.226, and 0.0216.}
\label{Fig8}
\end{figure}

\begin{figure}[t]
\centerline{\includegraphics[width=7cm]{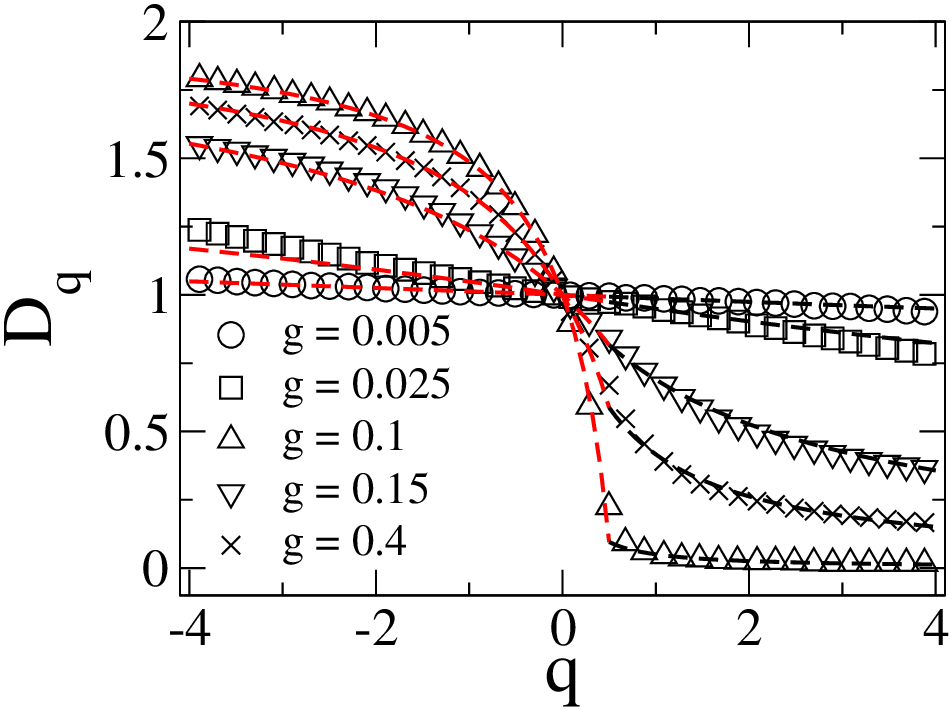}}
\caption{(Color online) $D_q$ as a function of $q$ for the CMT ensemble for several values of $g$.
The data was taken from Ref.~[\onlinecite{BG11c}]. Black [Red] dashed lines are Eq.~(\ref{DqofD1}) [Eq.~(\ref{D1_qofD1})].
We used $D_1=0.987$, 0.949, 0.836, 0.416 and 0.0502.}
\label{Fig9}
\end{figure}

In Figs.~\ref{Fig8} and \ref{Fig9} we plot $D_q$ as a function of $q$ for the CMR and CMT ensembles, respectively, for several values of $g$.
To have an independent verification of our predictions, the data reported in these figures was taken from Ref.~[\onlinecite{BG11c}]. 
We compare the numerical data with our equations for $D_q$ with $q<1/2$, Eq.~(\ref{D1_qofD1}), and $q>1/2$, Eq.~(\ref{DqofD1}); using as input,
values of $D_1$ obtained by the interpolation of the curves $D_q$ vs. $q$. We observe for both ensembles that our predictions reproduce reasonably well 
the numerical data. 

\subsection{The Ruijsenaars-Schneider Ensemble and Intermediate quantum maps}

The Ruijsenaars-Schneider Ensemble (RSE) proposed in [\onlinecite{BGS09}] is defined as matrices of the form
\begin{equation}
    H_{mn} = \exp(i\Phi_m) \frac{1-\exp(2\pi ig)}{N[1-\exp(2\pi i(m-n+g)/N)]} \ ,
\label{RSE}
\end{equation}
where $1\le m\le n$, $\Phi_m$ are independent random phases distributed between 0 and $2\pi$, and $g$ is a free parameter independent on $N$.

\begin{figure}[t]
\centerline{\includegraphics[width=7cm]{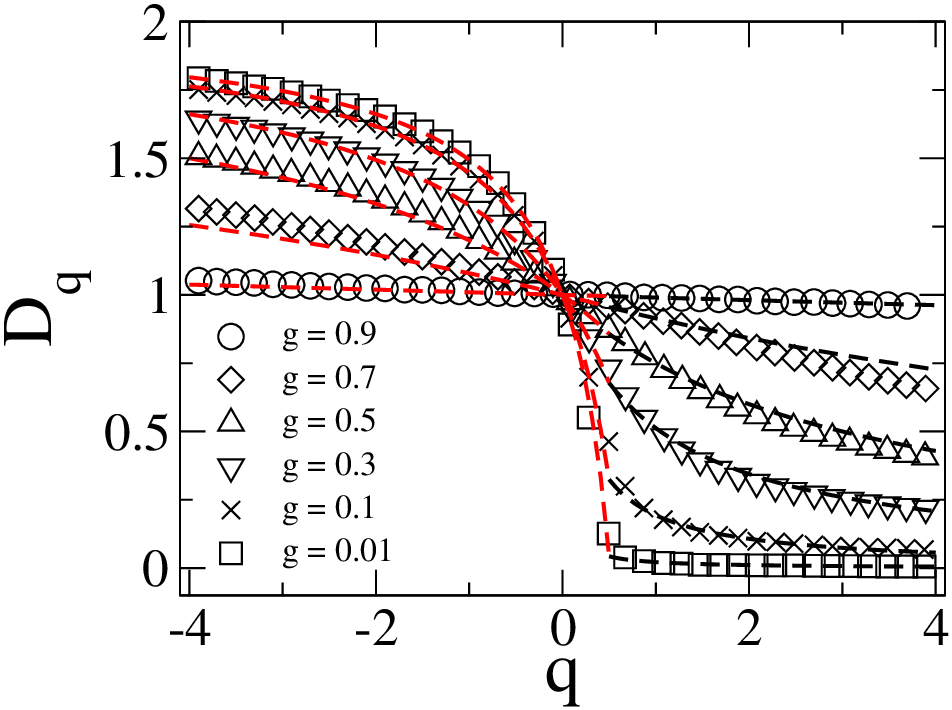}}
\caption{(Color online) $D_q$ as a function of $q$ for the RSE for several values of $g$.
The data was taken from Ref.~[\onlinecite{BG11c}]. Black [Red] dashed lines are Eq.~(\ref{DqofD1}) [Eq.~(\ref{D1_qofD1})].
We used $D_1=0.99$, 0.914, 0.75, 0.512, 0.193, and 0.022.}
\label{Fig10}
\end{figure}

Now, in Fig.~\ref{Fig10} we present $D_q$ as a function of $q$ for the RSE for several values of $g$. The data was also taken from 
Ref.~[\onlinecite{BG11c}]. As for the CM ensembles, here we observe that our predictions reproduce reasonably well the numerical data for both $D_q$ with $q<1/2$
and $q>1/2$. The values of $D_1$ we used as input in Eqs.~(\ref{D1_qofD1}) and (\ref{DqofD1}) were obtained by the interpolation of the curves $D_q$ vs. $q$.

In fact, in Ref.~[\onlinecite{EPL12}] we have already successfully tested the predictions given by Eq.~(\ref{D2ofD1}) on the RSE. There, we even formulated
heuristic expressions for $D_q$:
\begin{equation}
     D_q \approx \left[ 1-(g-1)^2 \right] \left[1+(q-1)(g-1)^2 \right]^{-1}
\label{DqRSE2}
\end{equation}
and
\begin{equation}
    D_q \approx \left[ k^2-(g-k)^2 \right] \left[k^2+(q-1)(g-k)^2 \right]^{-1} \ .
\label{DqRSE2k}
\end{equation}
for $0<g<1$ and $|g-k|\ll 1$ with $k\ge 2$, respectively. We got the expressions above by substituting $\chi \sim (g-1)^2$ and
$\chi \sim (g-k)^2/k^2$ [or $D_1 \approx 1-(g-1)^2$ and $D_1 \approx 1-(g-k)^2/k^2$], obtained analytically in Ref.~[\onlinecite{BG11}], into Eq.~(\ref{Dqofchi}) [or Eq.~(\ref{DqofD1})].

Also, in Ref.~[\onlinecite{EPL12}] we tested some of our predictions for a variant of the RSE, introduced in [\onlinecite{MGG08}], with the
name of intermediate quantum maps (IQM) model; see also [\onlinecite{MGGG10}]. In this model the parameter $g$ of the RSE is replaced by $cN/g$ with $cN=\pm 1$ mod $g$, 
being $g$ the parameter of the IQM model. For the IQM model we substituted $\chi \approx 1/g$ or $D_1\approx 1-1/g$, analytical expressions reported in [\onlinecite{MGG08}], 
into Eq.~(\ref{Dqofchi}) or (\ref{DqofD1}), respectively, to get
\begin{equation}
    D_q \approx \left( g-1 \right) \left( g+q-1 \right)^{-1} \ .
\label{DqIQM2}
\end{equation}
\begin{figure}[t]
\centerline{\includegraphics[width=7cm]{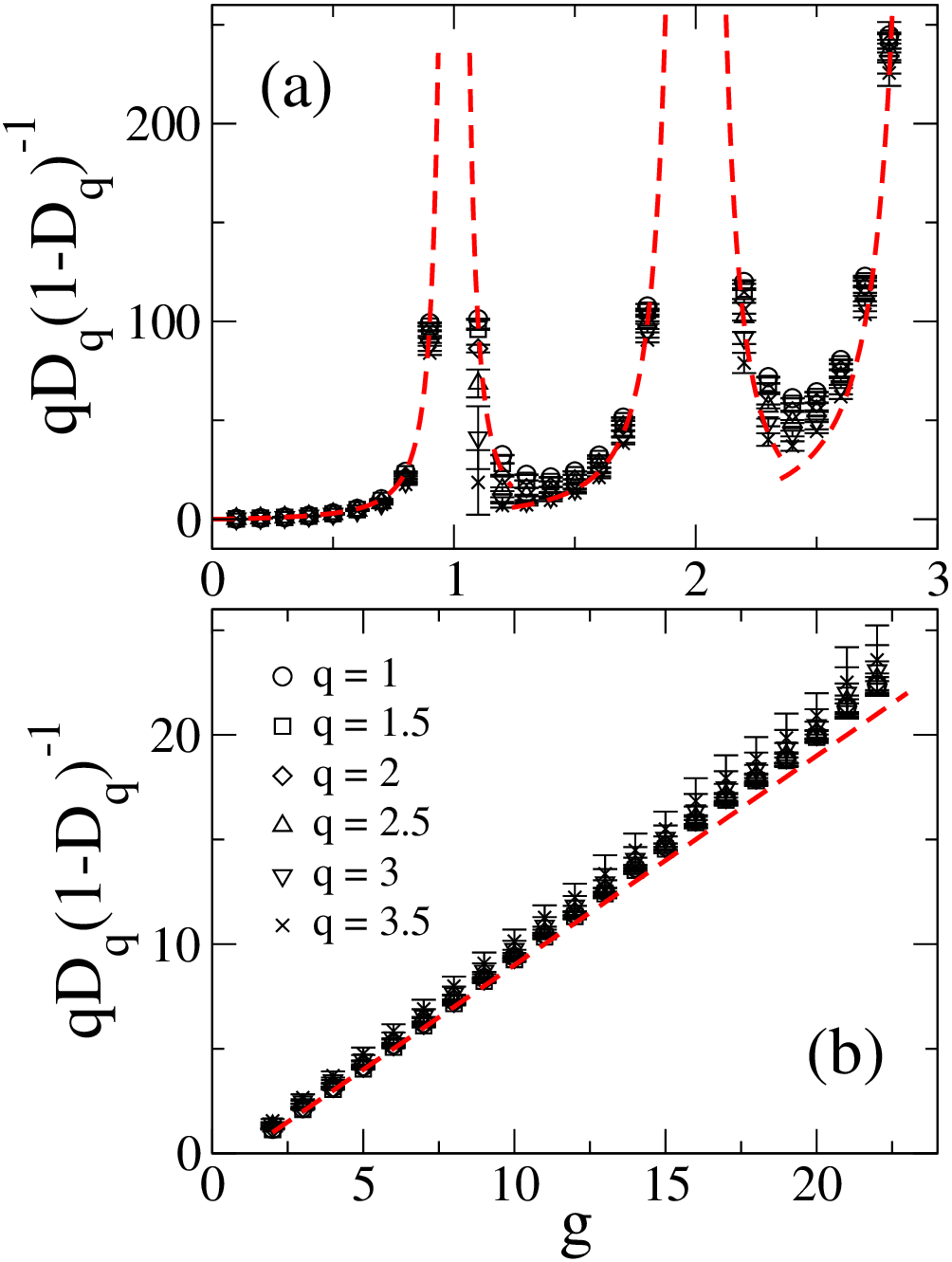}}
\caption{(Color online) $qD_q(1-D_q)^{-1}$ as a function of $g$ for (a) the RSE and (b) IQM model for several values of $q$. 
Red lines are (a) Eqs.~(\ref{RSEa}-\ref{RSEb}) and (b) Eq.~(\ref{IQMa}). The multifractal dimensions $D_q$ reported in this figure  
were computed using the same matrix sizes and ensemble realizations as for the PBRM model.}
\label{Fig11}
\end{figure}

Here we just want to add that by the use of Eqs.~(\ref{DqRSE2}) and (\ref{DqRSE2k}), for the RSE, and Eq.~(\ref{DqIQM2}), for the IQM model, we can demonstrate the
independence of $qD_q(1-D_q)^{-1}$ on $q$ (already shown for the PBRM model in Fig.~\ref{Fig3}). In fact, by substituting the above-mentioned expressions into Eq.~(\ref{DqpDq})
we get
\begin{equation}
    qD_q(1-D_q)^{-1} \approx (g-1)^{-2} - 1 \ ,
\label{RSEa}
\end{equation}
and
\begin{equation}
    qD_q(1-D_q)^{-1} \approx k^2(g-k)^{-2} - 1 \ ,
\label{RSEb}
\end{equation}
for $0<g<1$ and $|g-k|\ll 1$ with $k\ge 2$, respectively, for the RSE; and
\begin{equation}
    qD_q(1-D_q)^{-1} \approx g-1 \ ,
\label{IQMa}
\end{equation}
for the IQM model. Then, In Fig.~\ref{Fig11} we plot $qD_q(1-D_q)^{-1}$ for the RSE and the IQM 
model for several values of $q$. We also include the equations given above in red dashed lines. We observe a rather good correspondence between numerical 
data and Eqs.~(\ref{RSEa}-\ref{IQMa}).

\subsection{Higher dimensional models}

The generalization of Eqs.~(\ref{Dqofchi}-\ref{D2ofD1}) to higher dimensional systems ($d>1$) can be done if $D_q$ is replaced by 
$D_q/d$ in Eqs.~(\ref{Dqofchi}-\ref{D2ofD1}). Then, below we explore the applicability of our results to the Quantum Hall transition in $d=2$ and the Anderson transition in $d=3$.

\begin{figure}[t]
\centerline{\includegraphics[width=7cm]{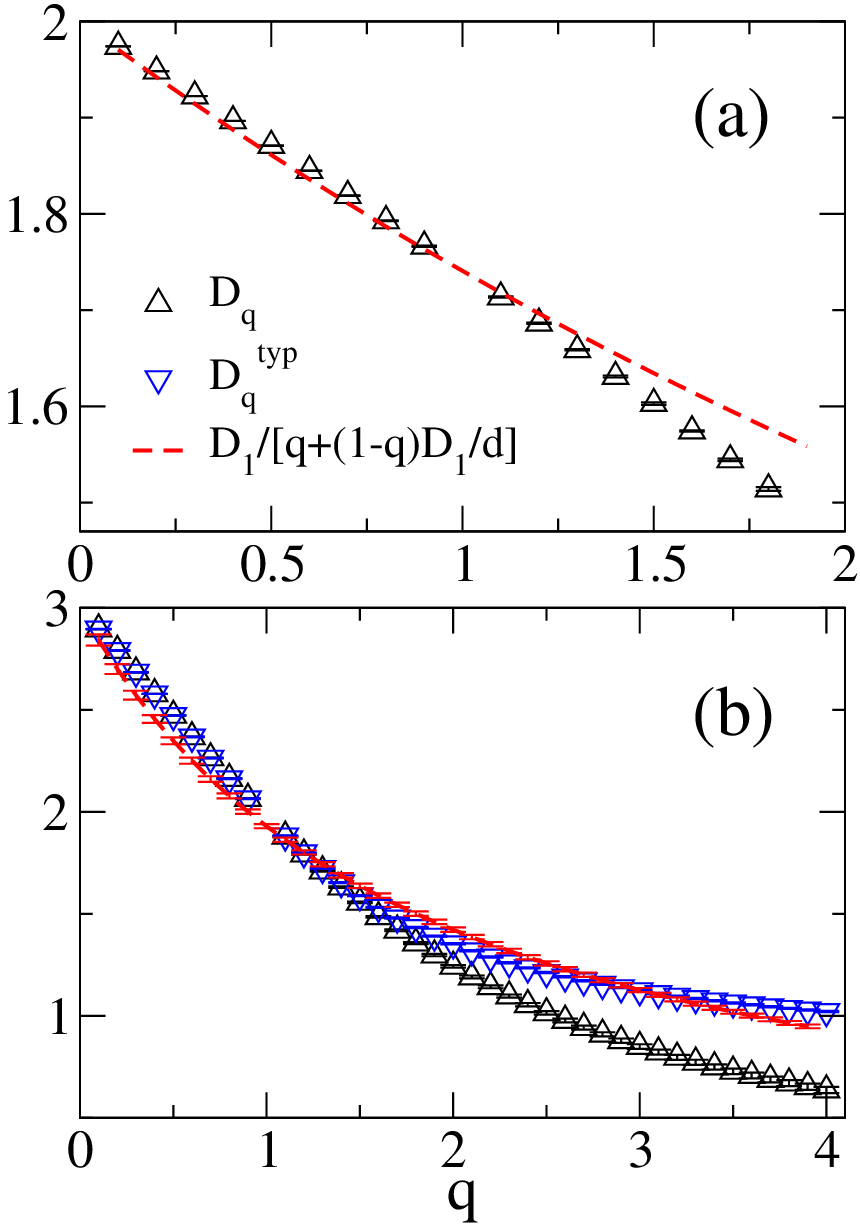}}
\caption{(Color online) (a) $D_q$ as a function of $q$ for the QHT. The red dashed 
line is the prediction for $D_q$ given by Eq.~(\ref{DqofD1}) using $D_1=1.7405\pm 0.0002$ [\onlinecite{EMM08}]. The numerical 
data for $D_q$ was taken from [\onlinecite{EMM08}]. (b)$D_q$ and $D_q^{\tbox{typ}}$ as a function of $q$ for the 3d Anderson model at criticality. 
The red dashed line is the prediction for $D_q$ given by Eq.~(\ref{DqofD1}) using $D_1=1.93\pm 0.01$ 
[\onlinecite{RVSR10}]. The numerical data for $D_q$ and $D_q^{\tbox{typ}}$ was taken from [\onlinecite{thesis}].}
\label{Fig12}
\end{figure}

In Fig.~\ref{Fig12}(a) we plot $D_q$ as a function of $q$ for the Quantum Hall transition (QHT). The data for $D_q$ was taken from [\onlinecite{EMM08}].
We also include the prediction for $D_q$ given by Eq.~(\ref{DqofD1}) (where $D_q$ has been replaced by $D_q/2$) using 
$D_1=1.7405\pm 0.0002$ [\onlinecite{EMM08}]. We observe that the prediction of Eq.~(\ref{DqofD1}) is a reasonably good approximation for $D_q$ in the interval $0<q<1.2$.

In Fig.~\ref{Fig12}(b) we plot $D_q$ and $D_q^{\tbox{typ}}$ as a function of $q$, for the 3d Anderson 
model at criticality, together with the prediction for $D_q$ given by Eq.~(\ref{DqofD1}) (where $D_q$ has been replaced by $D_q/3$) 
using $D_1=1.93\pm 0.01$ [\onlinecite{RVSR10}]. The data for $D_q$ and $D_q^{\tbox{typ}}$ was taken from [\onlinecite{thesis}].
The multifractal dimensions $D_q^{\tbox{typ}}$ were extracted from the scaling of the {\it typical} participation
numbers ${\cal I}_q^{\tbox{typ}}\equiv \exp\bra\ln{\cal I}_q\ket$ with the system size $N$, from the relation
\begin{equation}
{\cal I}_q^{\tbox{typ}}\sim N^{-(q-1)D_q^{\tbox{typ}}} \ .
\end{equation}

We observe that the prediction of Eq.~(\ref{DqofD1}) is reasonably good for $D_q^{\tbox{typ}}$ with $0<q<4$.
In contrast, Eq.~(\ref{DqofD1}) do not reproduce the numerical $D_q$ when $q>1$. We have also substituted $D_1=1.97\pm 0.002$ (obtained 
from the interpolation of the $D_q$ data from [\onlinecite{thesis}]) into Eq.~(\ref{DqofD1}) but the resulting $D_q$ curve is very similar to that with 
$D_1=1.93\pm 0.01$, so we do not show it in Fig.~\ref{Fig12}(b). 

\section{Applicability to deterministic models}

In the previous section we have verified that relations (\ref{Dqofchi}-\ref{D2ofD1}) are valid for critical random matrix 
ensembles in 1d and also, to some extent, to higher dimensional models at criticality. The common feature in the systems used above 
is the presence of multifractal eigenstates. However, note that not only disordered models produce them. It is well known that deterministic models having self-similar 
potentials also possess multifractal eigenstates, see for example [\onlinecite{FKT89,WS12}]. Moreover, multifractal 
objects can be expressly constructed by the use of iteration algorithms. As examples we can mention the Cantor set and the set produced by baker's map. 

Below, we test the applicability of our expressions relating multifractal exponents now to the multifractal eigenstates of
a tight-binding model having a self-similar potential and to multifractal sets produced by iteration algorithms.

\subsection{Off--diagonal one dimensional Fibonacci lattice}

According to Fujiwara {\it et al.} [\onlinecite{FKT89}] the multifractal spectrum of a one--dimensional Fibonacci sequence 
can be represented by the inflation rule $T_{n+1}=T_nT_{n-1}$, where $T_1={A}$ and $T_2={AB}$, so $T_3={ABA}$ and so on. In this case the Schr\"odinger equation
\[
    t_{j+1}\psi_{j+1}+t_{j-1}\psi_{j-1}=E\psi_j 
\]
has a multifractal solution at the bandcenter, $E=0$. Then, by defining the parameter $g=t_{AB}/t_{AA}$ the generalized dimensions of the eigenstates
take the form \cite{FKT89}
\begin{equation}
   D_q = (3\ln\sigma)^{-1}(q-1)^{-1} \ln\left[ \lambda^q(g^2)/\lambda (g^{2q}) \right] \ ,
\label{DqFibonacci}
\end{equation}
where $\sigma =(\sqrt{5}+1)/2$ is the golden mean, and $\lambda(x) = (2x)^{-1}[(x+1)^2+\sqrt{(x+1)^4+4x^2}]$.
Hence, the information dimension and the correlation dimension read as
\begin{equation}
   D_1=(3\ln\sigma)^{-1} \left [ \ln\lambda(g^2)-g^2\ln g^2 \lambda'(g^2)/\lambda(g^2) \right ]
\label{D1Fibonacci}
\end{equation}
and 
\begin{equation}
   D_2=(3\ln\sigma)^{-1} \left [ 2\ln\lambda(g^2)-\ln\lambda(g^4) \right ] \ ,
\end{equation}
respectively.

\begin{figure}[t]
\centerline{\includegraphics[width=8cm]{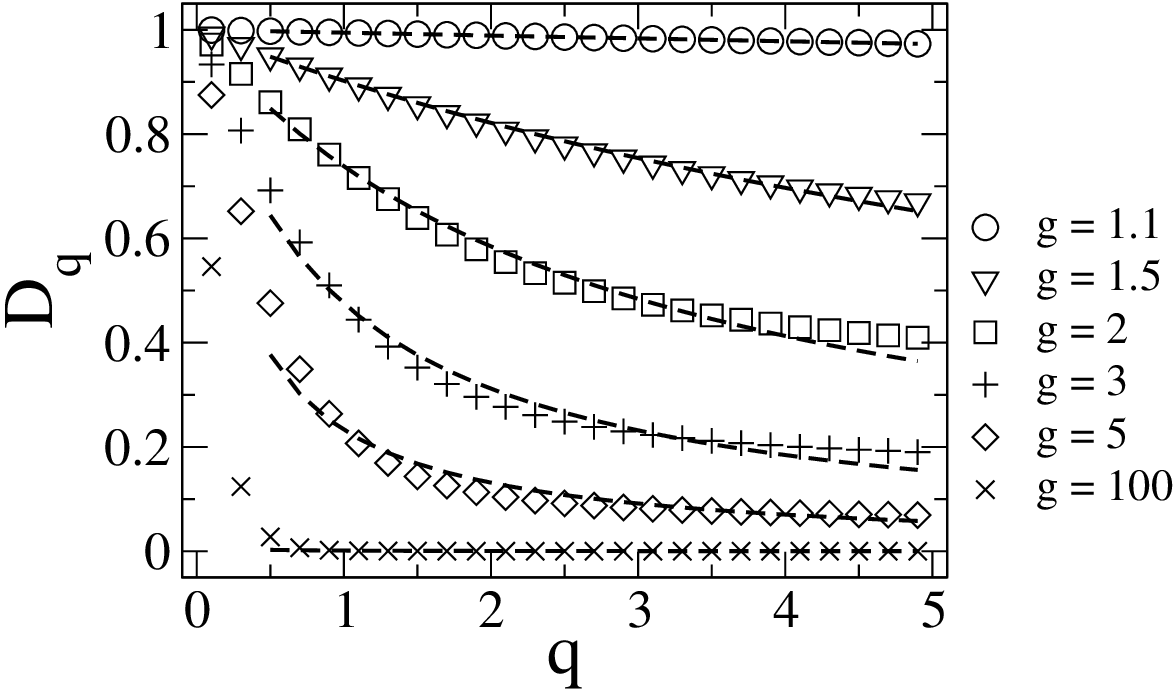}}
\caption{$D_q$, computed from Eq.~(\ref{DqFibonacci}), as a function of $q$ for the Fibonacci lattice for several values of $g$. The 
dashed line is the prediction for $D_q$ given by Eq.~(\ref{DqofD1}) with $D_1$ calculated from Eq.~(\ref{D1Fibonacci}).}
\label{Fig13}
\end{figure}

Then, in Fig.~\ref{Fig13} we plot $D_q$, computed from Eq.~(\ref{DqFibonacci}), as a function of $q$ for the Fibonacci 
lattice for several values of $g$. The dashed line is the prediction for $D_q$ given by Eq.~(\ref{DqofD1}) 
with $D_1$ calculated from Eq.~(\ref{D1Fibonacci}). Again, as for the eigenstates of disordered models, here we observe that 
Eq.~(\ref{DqofD1}) reproduces rather well the multifractal dimensions of the eigenstates of the Fibonacci lattice, mainly in the range of $1<q<4$.

\subsection{The 2--measure, 1--scale Cantor set or binomial branching process}

In order to generate a multifractal distribution we use a Cantor set with two measures.\cite{Cantor} Take the $[0,1]$ 
interval and divide it into two parts, which in our case can cover the whole interval, and even we can choose them to be equal. 
So let us partition the unit interval into two equal halves. Now let us introduce another measure: let us associate the 
probability $g$ with one of the intervals and $(1-g)$ with the other one. Then let us do the same procedure with the two 
subintervals. In this way we start with the unit interval with probability measure 1 in the 0th approximation, then the 1st 
iteration gives two half intervals with probabilities $g$ and $1-g$. After the 2nd iteration we get four intervals, with length 
one-fourth each but with measures $g^2$, $g(1-g)$, $(1-g)g$, and $(1-g)^2$. Proceeding further iteratively the distribution 
will be a multifractal. Such a distribution may be achieved also using a binomial branching process.\cite{MH13}

Since the distribution can be obtained in a recursive way, the generalized entropies of these distributions can be traced back
to the 1st iteration, so we can write down $D_q$ immediately, see [\onlinecite{VPJP96,MH13}] for details. Therefore,
\begin{equation}
    D_q = \ln \left[g^q + (1-g)^q \right] [(1-q)\ln 2]^{-1}
\label{DqCantor}
\end{equation}
and
\begin{equation}
    D_1 = -\left[ g\ln g + (1-g)\ln (1-g) \right ] (\ln 2)^{-1} \ ,
\label{D1Cantor}
\end{equation}
where the factor $\ln 2$ comes from the fact that the unit interval has been divided into two pieces of length $1/2$ or in other words 
the branching is always two-fold.

\begin{figure}[t]
\centerline{\includegraphics[width=8cm]{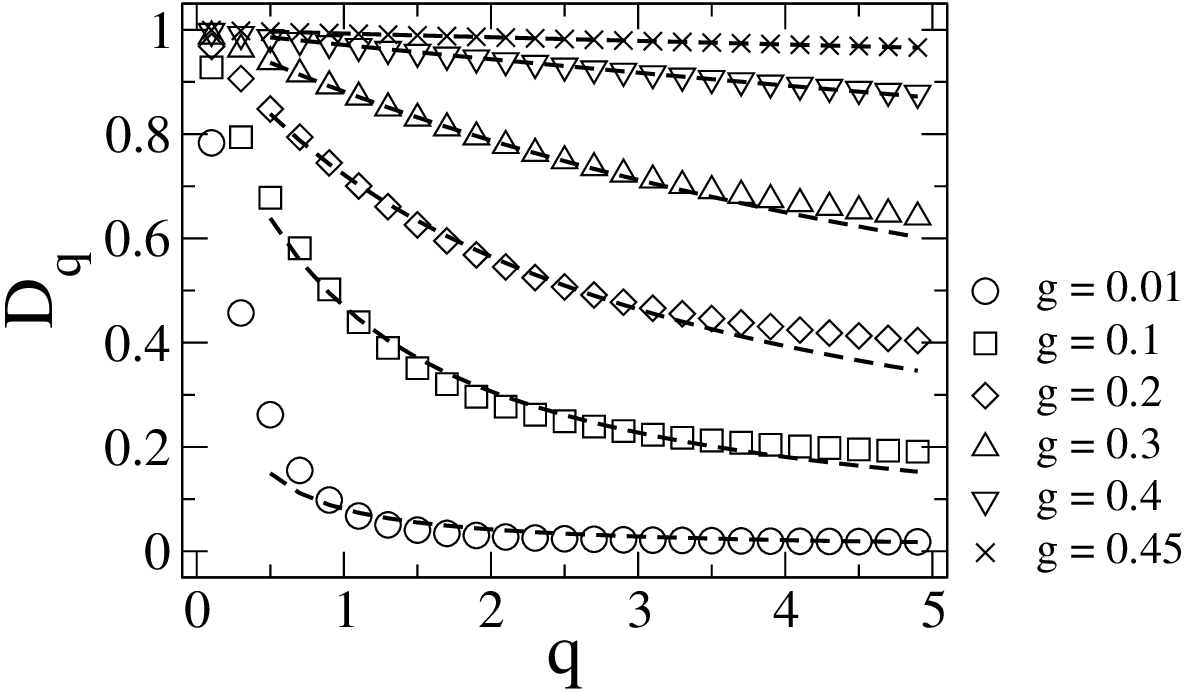}}
\caption{$D_q$, computed from Eq.~(\ref{DqCantor}), as a function of $q$ for the Cantor set for several values of $g$. The dashed 
line is the prediction for $D_q$ given by Eq.~(\ref{DqofD1}) with $D_1$ calculated from Eq.~(\ref{D1Cantor}).}
\label{Fig14}
\end{figure}

In Fig.~\ref{Fig14} we plot $D_q$, computed from Eq.~(\ref{DqCantor}), as a function of $q$ for the Cantor set for several values of $g$. 
The dashed line is the prediction for $D_q$ given by Eq.~(\ref{DqofD1}) with $D_1$ calculated from Eq.~(\ref{D1Cantor}). Again, as for the 
Fibonacci lattice, we observe that Eq.~(\ref{DqofD1}) reproduces rather well the multifractal dimensions of the the Cantor set, mainly in the range of $1<q<4$.
\begin{figure}[t]
\centerline{\includegraphics[width=8cm]{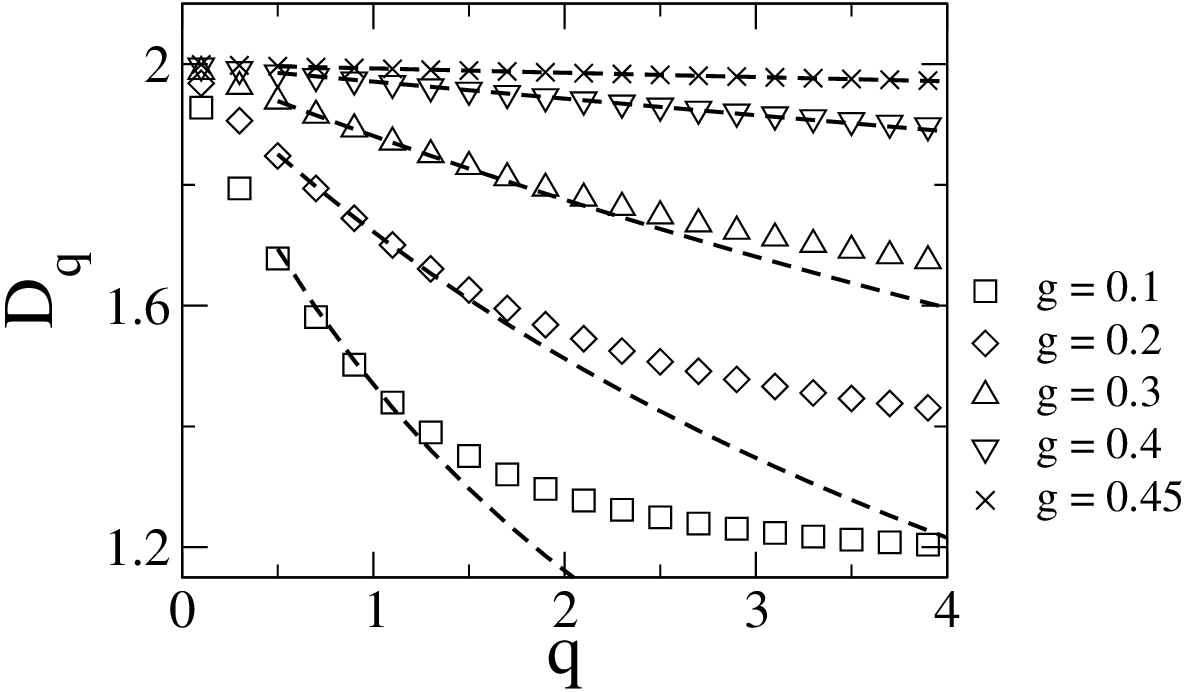}}
\caption{$D_q$, computed from Eq.~(\ref{DqBaker}), as a function of $q$ for the generalized baker's map for several values of $g$ 
and $\lambda=0.5$. The dashed line is the prediction for $D_q$ given by Eq.~(\ref{DqofD1}) with $D_1$ extracted from the interpolation of the data $D_q$ vs. $q$.}
\label{Fig15}
\end{figure}

\subsection{Generalized baker's map}

The generalized baker's map is defined as a transformation of the unit square $[0,1]\times[0,1]$ with the following rules:\cite{Ott} We first 
divide the unit square into two pieces, $y<g$ and $y>g$; $g=(0,1)$. We then compress the two pieces in the horizontal direction by different 
factors, $\lambda_a$ for the piece in $y<g$ and $\lambda_b$ for the piece in $y>g$; where $\lambda_a+\lambda_b\le 1$. Then we vertically
stretch the lower piece by a factor $1/g$ and the upper piece by a factor $1/(1-g)$, so that both are of unit length. We then take the upper piece
and place it back in the unit square with its right vertical edge coincident with the right vertical edge of the unit square. Thus, the 
generalized baker's map is a mapping of the unit square into two stripes within the square, one in $0\le x \le\lambda_a$ and another one in 
$1-\lambda_b\le x \le 1$. Applying the map a second time, maps the two stripes into four stripes. Application of the map $n$ more times, 
results in more stripes of narrower width, where the widths approach zero as $n$ approaches infinity. In fact, the intersection of the attractor 
with a horizontal line is a Cantor set. In the particular case of $\lambda=\lambda_a=\lambda_b$ we have\cite{Ott}
\begin{equation}
     D_q = 1 + \frac{1}{q-1} \frac{\ln(g^q + (1-g)^q)}{\ln(\lambda)} \ .
\label{DqBaker}
\end{equation}

Finally, in Fig.~\ref{Fig15} we plot $D_q$, computed from Eq.~(\ref{DqBaker}), as a function of $q$ for the generalized baker's map for several values of $g$. 
The dashed line is the prediction for $D_q$ given by Eq.~(\ref{DqofD1}) with $D_1$ calculated from the interpolation of the data $D_q$ vs. $q$. 
Note that this multifractal is embedded in 2d, so we make the substitution $D_q\to D_q/2$ in Eq.~(\ref{DqofD1}). We observe that Eq.~(\ref{DqofD1}) reproduces 
well the multifractal dimensions of the generalized baker's map mainly when $g\to 1/2$. When, $g\to 0$ we observe good correspondence in the range $1/2<q<1$, only.

\section{Conclusions}

In this paper we propose heuristic relations on one hand between the generalized multifractal dimensions, $D_q$ and $D_{q'}$, for a relatively wide range of the
parameter $q$, and on the other hand between these dimensions and the level compressibility $\chi$. As a result we find a general framework embracing an 
earlier result\cite{CKL96} and a recent one.\cite{BG11} Our proposed relations have been corroborated by numerical simulations on: 
various random matrix ensembles, a deterministic model having a self-similar potential whose eigenstates have 
multifractal properties, and multifractal objects expressly constructed by the use of iteration algorithms. Of course the analytical relations
and the numerical simulations set limitations on the validity for a certain range of the parameter $q$. Therefore our results are obviously approximate. Hence they 
call for further theoretical as well as numerical investigations.

Moreover, since our relations between the generalized dimensions and the level compressibility allowed us to state a clear link between the spectral and 
scattering properties of disordered systems at the metal-insulator transition, it may be interesting to explore the consequences of our results on the
quantities characterizing the dynamical properties of critical random matrix ensembles; which have been the focus of very recent investigations.\cite{KOY11,KYSC12,GMGG12}

We believe that our results may find applications in several recently studied models characterized by multifractal eigenstates; such as deterministic 
self-similar potentials,\cite{FKT89,WS12} quantum spin chains,\cite{AB12} Dirac fermions in the presence of random magnetic fields,\cite{CHHF12} 
disordered graphene,\cite{graphene} and other critical random matrix ensembles.\cite{BG11,FOR09}

\begin{acknowledgments}
The authors are greatly indebted to V. Kravtsov for useful discussions. This work was partially supported by VIEP-BUAP (Grant MEBJ-EXC14-I), 
PIFCA (Grant BUAP-CA-169), the Alexander von Humboldt Foundation, and the Hungarian Research Fund (OTKA) grant K108676.
\end{acknowledgments}

\end{document}